%% file: Karachentsev_eng.tex
\documentclass[
aps,%
12pt,%
final,%
notitlepage,%
oneside,%
onecolumn,%
nobibnotes,%
nofootinbib,%
superscriptaddress,%
noshowpacs,%
centertags]%
{revtex4}
\usepackage[koi8-r]{inputenc}
\usepackage{graphicx}
\usepackage[russian]{babel}

\begin{document}

\title{Optical Identifications of 230 HIPASS Radio Sources}

\author{\firstname{I.~D.}~\surname{Karachentsev}}
\affiliation{Special Astrophysical Observatory, Russian Academy of Sciences, Nizhnii Arkhyz, 357169 Karachaevo-Cherkesskaya Republic, Russia}
\email{ikar@sao.ru}
\author{\firstname{D.~I.}~\surname{Makarov}}
\affiliation{Special Astrophysical Observatory, Russian Academy of Sciences, Nizhnii Arkhyz, 357169 Karachaevo-Cherkesskaya Republic, Russia}
\affiliation{Invited researcher in Lyon Observatory, Lyon, France}
\author{\firstname{V.~E.}~\surname{Karachentseva}}
\author{\firstname{O.~V.}~\surname{Melnyk}}
\affiliation{Astronomical Observatory, Kiev National University, Observatorna vul. 3, Kiev, 04053 Ukraine}

\today
\begin{abstract}
We present the coordinates, apparent magnitudes, and morphological types for 230 galaxies
presumably identified with HIPASS (HI Parkes All-Sky Survey) sources. The new optical counterparts
of the HIPASS sources follow the well-known statistical relationships between the hydrogen mass,
luminosity, and type of galaxies. Low-surface-brightness galaxies constitute a significant fraction among
these objects. The median value of the hydrogen mass-to-luminosity ratio for them is a factor of 2 or 3
higher than that for bright HIPASS galaxies, reaching $1.7 M_{\odot}/L_{\odot}$. A number of our objects are located
near the boundary $\log(M_{HI}/L_B) = 0.2(M_B + 20)$ that defines the zone of gravitational stability of disk
galaxies against large-scale star formation.

\end{abstract}

\maketitle

\section{INTRODUCTION}

The blind survey of the southern sky in the 21-cm
HI line carried out with the 64-m radio telescope in
Parkes (Australia) has revealed 4315 radio sources.
The catalogs of HIPASS sources were published by
Meyer et al. (2004), Koribalski et al. (2004), and Henning
et al. (2000). The effective depth of this survey
in radial velocities is $\sim3000$ km s$^{-1}$, i.e., most of
the HIPASS sources are located in the Local Supercluster.

Numerous optical identifications of HIPASS objects
were made by Staveley-Smith et al. (1998),
Ryan-Weber et al. (2002), and Doyle et al. (2005).
The latest list of optical identifications (http://hipass.aus-vo.org) contains probable optical counterparts
(galaxies) for 3618 HIPASS sources, i.e.,
84\% of their total number. The remaining 16\%
of the objects have no optical identifications for
various reasons: (a) two or more galaxies fall into
the wide $(\sim15^{\prime})$ beam of the radio telescope; (b)
objects are located at low Galactic latitudes in the
zone of strong absorption; (c) a HIPASS source is a
dwarf galaxy rich in gas but with a very low surface
brightness similar to those detected by Karachentseva
and Karanchentsev (1998), KK, Karachentseva
and Karanchentsev (2000), KKs, and Karachentsev
et al. (2000), KKSG, during an all-sky survey on
POSS-II images is a HIPASS source. Recently,
Wong et al. (2006) published a northern extension of
the Parkes Survey to a declination of $+25.5^{\circ}$, bringing
the total number of HIPASS radio sources to 5317.
For some of the northern HIPASS sources, theNASA
Extragalactic Database (NED) gives identifications
without any reference to their origin.

When preparing the list of galaxies within the
Local Supercluster (with radial velocities up to
3800 km s$^{-1}$), we paid attention to the unidentified
HIPASS sources. To supplement our list, we made an
independent optical check for more than 500 HIPASS
sources, which added over 200 more identifications.
In this paper, we present data for 230 identified
HIPASS sources, leaving without comments the
cases where no optical counterparts were found.

\section{RESULTS}

When examining the Digitized Sky Survey areas
with unidentified HIPASS sources, we took into account
the intensity and width of the 21-cm line as
well as the position of a particular galaxy relative to
the position of the HIPASS object. We excluded the
cases of a possible identification with several suitable
optical galaxies. For each galaxy, we determined
its morphological type T by taking into account its
B and R brightness and visibility in the 2MASS
survey as well as its total apparent B magnitude
from comparison with the images of other galaxies
of a similar structure with known magnitudes.
Occasionally, we included galaxies with previously
made identifications in our list by adding our own
estimates of the morphological type and magnitude.
Our identifications of 230 HIPASS sources are given
in the table\footnote{The table is published completely only in electronic form and is accessible via ftp 
cdsarc.u-strasbg.fr/pub/cats/J (130.79.128.5) or http://cdsweb.u-strasbg.fr/pub/cats/J.}. 
The columns of the table contain the
following data: (1) HIPASS source name; in several
cases, we added objects from the HIZOA (Donley
et al. 2005), HIDEEP (Minchin et al. 2003), and
HIJASS (Lang et al. 2003) surveys; (2) coordinates of
the optical galaxy for the epoch 2000.0; (3) heliocentric
radial velocities from the 21-cm line in km s$^{-1}$;
(4, 5) full width at half maximum (FWHM) of the
21-cm line, in km s$^{-1}$, and flux F in the 21 cm
line, in Jansky km s$^{-1}$, taken mainly from HOPCAT
(Doyle et al. 2005); (6) total apparent magnitude from
NED (given with an accuracy of 0.01$^m$) or from our
estimates with an accuracy of about 0.5$^m$; (7) galactic
extinction in the B band from Schlegel et al. (1998);
(8) hydrogen mass-to-luminosity ratio in units of the
solar mass and luminosity, which does not depend
on the galaxy distance; (9) morphological type in the
de Vaucouleurs numerical scale; (10) notes with a
description of structural features. In the notes, we also
give the galaxy names from NED (from the catalogs
and lists); in the references to the 2dF, 6dF, APM,
2MASS, and other surveys, the galaxy names are
given without any coordinate part.

Figure 1 presents negative DSS images of the
four galaxies that we identified with HIPASS sources.
The galaxies with a low surface brightness are at the
center of each image; the image size is 6$\times$6 arcmin.

\section{DISCUSSION}

Late-type and low-surface-brightness galaxies
prevail among our identifications as ``residual'' with
respect to the previous ones. About a quarter of the
galaxies are located at low Galactic latitudes with
extinction $A_b>1.0^m$. Figure 2 shows the distribution
of our identified galaxies in hydrogen mass-toluminosity
ratio on a logarithmic scale. The $M_{HI}/L$
distribution has a median of $1.7 M_{\odot}/L_{\odot}$; half of the
entire sample is contained between the quartile values
$(0.9\div2.4) M_{\odot}/L_{\odot}$. For comparison, note that the
sample of 752 HIPASS BGC galaxies (Koribalski
et al. 2004) has a median $M_{HI}/L$ that is two and a
half times smaller than that for the galaxies from the
table. In Fig. 3, the $M_{HI}/L$ ratio is plotted against
the morphological type of the galaxies identified in
our survey. As we see, more than half of the galaxies
belong to the latest types: 10 (Ir), 9 (Im, BCD), and
7, 8 (Sd, Sdm). An expected positive correlation with
the coefficient $R = 0.40$ and standard deviation 0.34
is seen between $\lg(M_{HI}/L)$ and $T$.

In Fig. 4, the hydrogen mass-to-luminosity ratio
is plotted against the observed hydrogen line FWHM
$W_{50}$ (Wong et al. 2006) (uncorrected for the slope).
As we see from the figure, the sources that we identified
confirm the well-known fact of a decrease in
the hydrogen mass fraction from faint (with smaller
$W_{50}$) to bright galaxies. Note that all galaxies with
$M_{HI}/L>5$ lie in the range $W_{50}<150$ km s$^{-1}$.

Another expected relationship between
$\lg(M_{HI}/L)$ and the absolute magnitude of a galaxy
MB calculated from its radial velocity with $H_0=72$ km s$^{-1}$ Mpc$^{-1}$ is shown in Fig. 5. The solid line
corresponds to the regression
$$ \lg(M_{HI}/L) = 0.15(M_B+17.5), \eqno(1)$$
which reflects the well-known statistical relationships
between the hydrogen mass of a galaxy and its
radius, $M_{HI}\propto R^2$, and between the total luminosity
and radius, $L\propto R^3$. Hence follows the relationship
$M_{HI}/L\propto L^{-1/3}$, which is almost equivalent to (1).
As we see from Fig. 5, there is no galaxy with a
$M_{HI}/L$ ratio smaller than $3\cdot M_{\odot}/L_{\odot}$ among the
dwarf galaxies fainter than $B_T$. Of course, the low
accuracy of the BT estimate from DSS images introduces
an additional scatter on the diagrams.

Warren et al. (2007) also constructed the relationship
$M_{HI}\propto M_B$ for HIPASS galaxies and discussed
it in connection with the problem of the minimum
number of stars that can be formed in a dwarf system
if it is stable. Considering HIPASS galaxies and
invoking four known objects with detailed photometry
and record high values of $M_{HI}/L>5M_{\odot}/L_{\odot}$,
(ESO~215-09, DDO~154, UGCA~292, NGC~3741),
these authors found a positive correlation between
$M_{HI}/L$ and the luminosity of the galaxies that is
almost identical to ours. Warren et al. (2007) pointed
out the presence of an upper limit on this diagram,
$$\lg(M_{HI}/L)_{\max} =0.20(M_B+20),\eqno(2)$$
under which the entire set of galaxies known to date
lies. The authors explain the presence of this empirical
limit as follows. According to Taylor and Webster
(2005), the galactic gas disk in the absence of an
internal stellar radiation field becomes gravothermally
unstable, which gives rise tomolecular hydrogen even
for galaxies with very low baryon masses. The fraction
of the unstable gas decreases from normal to progressively
less massive galaxies. As a result, dwarf
galaxies become stable at a smaller fraction of the
gas transformed into stars. Thus, the presence of the
upper limit (2) can be interpreted as the existence of
a minimum fraction of the baryon mass in a galaxy
that must be transformed into stars for the galaxy to
remain stable against large-scale star formation.

We can add four more galaxies from the Catalog of
Nearby Galaxies (Karachentsev et al. 2004) with high
values of $(M_{HI}/L)$ to the four mentioned galaxies
with record high ratios $M_{HI}/L$ (ESO~215--09 (22),
DDO~154 (9.4),  UGCA~292 (7.0), and
NGC~3741 (5.8)): HolmIX (11.0), ESO~564--030 (6.3), PGC~51659 (6.0), and AndIV (21.0). The
hydrogen data for AndIV were taken from Chengalur et al. (2007). The hydrogen cloud in the Virgo Cluster
(HI~1225+01) discovered by Giovannelli and Haynes
(1989) with signatures of recent star formation
(Salzer 1992) is also plotted on the diagram of Fig. 5.
Based on the results of our work, we found ten more
galaxies with a high gas content per unit luminosity:
J0718-09 (6.6), J0727+04 (7.4), J1007-66 (12.4),
Leib (19.1), KKSG~47 (7.8), J1549+16 (11.0),
J1739-51 (10.4), J2011-09 (5.4), J2015+12 (6.6),
and ESO~403--036 (5.1). Among them, only one
galaxy, J1739-51, exceeds significantly the critical
condition (2). However, its visual $B_T$ estimate needs
a photometric improvement.

Note that our sample includes the unusual dwarf
system HIPASS J1123+13 in the nearby group
around NGC~3628. We identified this HIPASS
source with an object of an extremely low surface
brightness with angular sizes of 1.4$^{\prime}\times 0.8^{\prime}$ and a
total apparent magnitude $B_T =17.5^m$. According to
Haynes et al. (1979), an HI filament, which is also
barely seen optically, extends from this source to the
neighboring bright galaxy NGC~3628. Obviously, this
exotic object, whichwe called ``Leib'', deserves a more
detailed study at various wavelengths.

\section{ACKNOWLEDGMENTS}

We wish to thank B. Koribalski and L.N. Makarova
for advice and help in the work. This work was
supported in part by theRussian Foundation for Basic
Research (project nos. 07--02--00005, 08--02--00627,
and 06--02--04017).

{}

\newpage

\input{Karachentsev_table_eng.tex}

\clearpage

\begin{figure}
\includegraphics[scale=0.75]{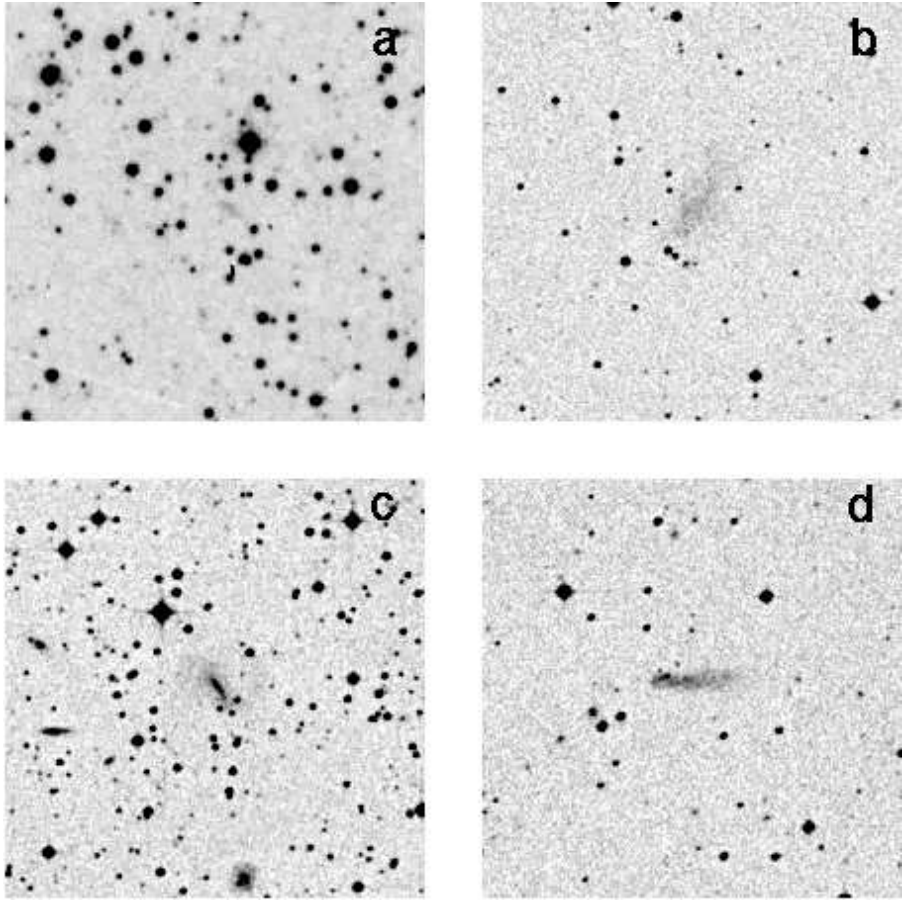}
\caption{
Digitized Sky Survey images of four galaxies identified with HIPASS sources: (a) HIPASS~J0727+04, (b)
HIPASS~J1435-17, (c) HIPASS~J2011-09, and (d) HIPASS~J2149-35. The size of each field is $6\times6$ arcmin; north is
at the top and east is to the left.}
\end{figure}

\begin{figure}
\caption{Distribution of the identified HIPASS sources in hydrogen mass-to-blue luminosity ratio in units of the solar mass and luminosity.}
\end{figure}

\begin{figure}
\caption{Hydrogen mass-to-luminosity ratio versus morphological type for 230 HIPASS sources.}
\end{figure}

\begin{figure}
\caption{Hydrogen mass-to-luminosity ratio versus observed line FWHM $W_{50}$ for 230 HIPASS sources.}
\end{figure}

\begin{figure}
\caption{Distribution of the identified HIPASS sources (circles) in absolute magnitude and hydrogen mass-to-luminosity
ratio. The statistical regression line (1) for them is indicated by the solid straight line. Its slope corresponds to the wellknown
correlations between the hydrogen mass, radius, and luminosity of disk galaxies. Other nearby galaxies with a high HI
abundance are indicated by the triangles. The dashed line indicates the minimum stellar mass fraction required to maintain the
gravitational stability of a galaxy according toWarren et al. (2007). The cross marks a typical error of our apparent magnitude
estimate for the galaxies.}
\end{figure}

\end{document}

%% file: Karachentsev_table_eng.tex
\clearpage
\hoffset=-2cm
\small
\begin{table}
\caption{New probable optical identifications of HIPASS sources in the Local Supercluster (N = 230).}
\begin{tabular}{lrrrrlrrrl} \hline

HIPASS (NED)     &  Optical counterpart& $V_h$ &   $ W_{50}$    & flux &   $B_t$    &  $A_b $&       $M_{HI}/L$ &$T$ &     Comments  \\
\hline
\multicolumn{1}{c}{(1)}&
\multicolumn{1}{c}{(2)}&(3)&(4)&(5)&(6)&(7)&(8)&(9)&(10)\\
\hline
HIPASS J0002$-$07  &  J000202.3$-$073913  &  3765   &   45   &  2.2  & 17.21  &  0.13  &  2.27   &    8& 6dF                       \\
HIPASS J0002$-$15  &  J000225.8$-$155327  &  3416   &   94   &  3.7  &  16.7  &  0.12  &  2.40   &    6&                           \\
HIPASS J0021$+$08  &  J002041.7$+$083656  &   693   &   41   &  1.7  &  16.8  &  0.55  &  0.82   &   10&                           \\
HIPASS J0030$-$52  &  J003028.6$-$521515  &  2640   &  151   &  2.6  & 16.63  &  0.06  &  1.67   &    2&                           \\
HIPASS J0041$-$01b &  J004139.7$-$020042  &  1949   &   76   &  3.7  & 16.89  &  0.12  &  2.86   &   10&                           \\
HIPASS J0116$-$63  &  J011650.0$-$633000  &  2306   &  130   &  2.5  &  16.7  &  0.09  &  1.66   &    7&                     \\
HIPASS J0207$-$06  &  J020724.4$-$061348  &  3819   &   82   &  3.4  &  17.2  &  0.11  &  3.53   &   10&                           \\
HIPASS J0227$-$15  &  J022720.8$-$152517  &  3767   &  254   & 10.3  & 14.85  &  0.09  &  1.25   &    7& MCG-03-07-19              \\
HIPASS J0237$+$12  &  J023718.6$+$123108  &   960   &   51   &  1.9  &  16.4  &  0.63  &  0.59   &    9&                           \\
HIPASS J0258$-$14  &  J025812.6$-$142302  &  3272   &  152   &  6.4  & 16.47  &  0.18  &  3.19   &    8& APMUKS                    \\
HIPASS J0305$-$69  &  J030525.2$-$685614  &  1369   &  120   &  7.3  &    16.0  &  0.15  &  2.42   &    9& bluish, IRAS               \\
HIPASS J0341$+$24  &  J034111.9$+$240046  &  1260   &  110   & 11.9  &  15.8  &  1.68  &  0.80   &    6&                           \\
HIPASS J0341$+$18  &  J034201.9$+$180831  &  1297   &   45   &    3.0  &  16.4  &  0.91  &  0.71   &    8&                           \\
HIPASS J0358$+$10  &  J035824.2$+$095845  &  1978   &  141   &  6.9  &    17.0  &  0.71  &  3.44   &    8&                           \\
HIPASS J0403$-$01  &  J040331.3$-$015540  &   937   &   92   &  9.5  &  16.4  &  0.93  &  2.21   &   10& LSB in a star halo        \\
HIPASS J0419$-$21  &  J042012.7$-$211436  &   900   &   46   &  2.7  &  16.3  &  0.12  &  1.21   &    8& ESO 550-23                \\
HIPASS J0421$-$51  &  J042036.0$-$511301  &  1046   &  142   &  5.9  &  16.5  &  0.06  &  3.37   &    9& LSBG F202-83?, 12'away     \\
HIPASS J0423$-$56  &  J042251.7$-$561332  &  1356   &   49   &  1.1  & 16.25  &  0.08  &  0.49   &   10& APM                       \\
HIPASS J0438$+$00  &  J043846.1$+$001845  &  3302   &   51   &  2.2  &  17.1  &  0.19  &  1.94   &   10&                           \\
HIPASS J0458$-$07  &  J045839.7$-$073306  &  4052   &   87   &  3.7  &    17.0  &  0.27  &  2.75   &    8&                           \\
HIPASS J0501$-$45  &  J050103.8$-$451450  &  1140   &   76   &  2.5  & 16.06  &  0.08  &  0.93   &    8& APM                       \\
HIPASS J0519$-$04  &  J051915.6$-$042608  &  3848   &   53   &  3.1  &  16.8  &  1.12  &  0.88   &    8&                           \\
HIPASS J0526$-$61  &  J052617.8$-$612141  &  1219   &  101   &  7.7  &  15.3  &  0.21  &  1.27   &    8&                           \\
HIJASS J0529$+$72  &  J052920.5$+$722727  &  1089   &   46   &   11.0  &    16.0  &  0.44  &  2.81   &   10& HS98                      \\
HIPASS J0531$+$08  &  J053102.4$+$082104  &   961   &   90   &  5.4  &  16.9  &  1.89  &  0.82   &   10&                           \\
HIPASS J0544$-$16  &  J054423.6$-$162652  &  2158   &  152   &  3.9  &  16.1  &  0.38  &  1.15   &    7&                           \\
HIPASS J0554$-$71  &  J055441.2$-$715539  &  1482   &   64   &  3.1  &  16.9  &  0.32  &  2.01   &    9&                           \\
HIPASS J0559$-$42  &  J055954.3$-$425314  &  1013   &  106   &  5.2  &  16.4  &  0.26  &  2.25   &   10&                           \\
\end{tabular}
\end{table}

\begin{table}
\begin{tabular}{lrrrrlrrrl} \hline
\hline
\multicolumn{1}{c}{(1)}&
\multicolumn{1}{c}{(2)}&(3)&(4)&(5)&(6)&(7)&(8)&(9)&(10)\\
\hline

HIPASS J0605$-$02  &  J060552.2$-$022216  &  2462   &  126   &  5.2  &  17.9  &   1.6  &  2.61   &   10&                           \\
HIPASS J0619$-$52  &  J061925.9$-$524638  &  1116   &   59   &  4.9  &  15.2  &  0.34  &  0.65   &   10&         \\
HIPASS J0620$-$27  &  J062007.1$-$272251  &  1789   &   47   &    3.0  &  16.8  &  0.17  &  2.03   &   10&                           \\
HIPASS J0620$+$20  &  J062103.5$+$201017  &  1318   &  138   &  8.7  &    18.0  &  3.47  &  0.86   &    7& ADBS (NED)                \\
HIPASS J0626$+$24  &  J062621.0$+$243920  &  1473   &   99   &  8.3  &  17.6  &  1.79  &  2.65   &    7&                           \\
HIPASS J0630$+$23  &  J062958.2$+$233428  &  1452   &  135   &   11.0  &  17.1  &  1.18  &  3.87   &    7&                           \\
HIPASS J0635$+$11  &  J063548.5$+$111507  &  3575   &   77   &    4.0  &  17.5  &  4.36  &  0.11   &    7&                           \\
HIPASS J0643$-$25  &  J064350.2$-$254509  &  2505   &  111   &    4.0  &  16.5  &  0.43  &  1.63   &    6&                           \\
HIPASS J0646$-$23  &  J064600.5$-$230144  &  1896   &   48   &  2.7  &    18.0  &  0.52  &  4.02   &   10&                           \\
HIPASS J0647$-$15  &  J064732.4$-$151611  &  2792   &  132   &  3.5  &  17.5  &  2.59  &  0.49   &    7&                           \\
HIPASS J0650$+$16  &  J065036.3$+$162121  &  2549   &   80   &  5.3  &  16.5  &  0.64  &  1.77   &    8&                           \\
HIPASS J0653$-$73  &  J065352.7$-$734228  &  1208   &   94   &  3.5  &    17.0  &  0.58  &  1.96   &    7&                           \\
HIPASS J0700$-$02  &  J070005.3$-$022352  &  1774   &   52   &  3.1  &  17.8  &  3.72  &  0.20   &   10&                           \\
HIPASS J0702$-$28  &  J070220.0$-$281859  &  1801   &   82   &  3.9  &    17.0  &  0.51  &  2.33   &   10& [DDR2005]     \\
HIZOA  J0702$+$03  &  J070250.7$+$031114  &  3551   &  103   &  4.2  &  18.2  &  1.52  &  2.99   &    8&                                                   \\
HIZOA  J0705$+$02  &  J070538.7$+$023720  &  1745   &   46   &  4.6  &  18.3  &     2.00  &  2.31   &    8& [H92]16ID                                         \\
HIPASS J0712$-$39  &  J071228.9$-$390549  &  2566   &   60   &  3.4  &  16.8  &  0.99  &  1.09   &    8&                                                   \\
HIPASS J0712$-$41  &  J071310.6$-$411935  &  1689   &   48   &  2.5  &  16.9  &  0.72  &  1.12   &    8&                                                   \\
HIPASS J0713$-$30  &  J071331.7$-$303537  &  1740   &   74   &    3.0  &    17.0  &  0.61  &  1.63   &    7&                                                   \\
HIPASS J0716$-$40b &  J071640.1$-$403305  &  3234   &   79   &  4.5  &  17.4  &  1.07  &  2.31   &    8&                                                   \\
HIPASS J0717$-$08  &  J071743.2$-$085517  &  2477   &  165   &  9.5  &  17.2  &  2.09  &  1.58   &    5&                                                   \\
HIPASS J0718$-$79  &  J071745.0$-$792440  &  1571   &   66   &  4.1  &  17.4  &  1.49  &  1.43   &   10&                                                   \\
HIPASS J0718$-$09  &  J071821.0$-$090316  &   909   &   57   & 13.2  &  18.2  &   1.90  &  6.61   &    8&                                                   \\
HIPASS J0721$-$30  &  J072106.2$-$300539  &  1951   &  105   &  6.6  &    17.0  &   0.90  &  2.75   &    7&                                                   \\
HIPASS J0725$-$24b &  J072520.6$-$245643  &  2769   &  232   & 12.1  &    18.0  &  4.59  &  0.42   &    6&                                                   \\
HIPASS J0727$+$04  &  J072754.0$+$044146  &  2087   &  121   &  4.5  &  17.9  &   0.30  &  7.45   &    7&                                                   \\
HIPASS J0737$-$31  &  J073717.1$-$310654  &  2237   &  157   &  7.7  &  17.6  &  2.46  &  1.33   &    6&                                                   \\
HIPASS J0741$-$09  &  J074134.4$-$094201  &  2435   &   96   &  2.7  &    17.0  &  0.76  &  1.28   &    6&                                                   \\
HIPASS J0742$-$26  &  J074158.5$-$264248  &  2924   &  237   &  8.1  &  17.7  &     4.00  &  0.37   &    6&                                                   \\
HIPASS J0745$-$74  &  J074521.0$-$744820  &  1741   &   90   &  1.8  &    18.0  &  2.01  &  0.68   &   10&                         \\
\end{tabular}
\end{table}

\begin{table}
\begin{tabular}{lrrrrlrrrl} \hline
\hline
\multicolumn{1}{c}{(1)}&
\multicolumn{1}{c}{(2)}&(3)&(4)&(5)&(6)&(7)&(8)&(9)&(10)\\
\hline

HIPASS J0747$-$26b &  J074712.8$-$261746  &  2580   &  170   &   19.0  &  17.1  &  2.54  &  1.92   &    6&                                                   \\
HIPASS J0751$-$37  &  J075136.9$-$371350  &  2808   &   87   & 10.3  &  18.2  &  5.26  &  0.23   &    7&                                                   \\
HIPASS J0752$-$49  &  J075218.7$-$493221  &  1158   &   56   &  2.6  &  16.7  &  0.89  &  0.83   &    8&                                                   \\
HIPASS J0754$-$38  &  J075448.4$-$385524  &  2757   &  270   & 16.4  &  17.2  &  3.47  &  0.77   &    7&                                                   \\
HIPASS J0758$+$10  &  J075812.3$+$110114  &  2346   &   94   &  5.7  &  16.4  &   0.10  &  2.86   &    7&                                                   \\
HIPASS J0759$-$71  &  J075923.5$-$715325  &  1368   &  110   &  2.4  &  17.5  &  1.11  &  1.31   &   10&                                     \\
HIPASS J0801$-$21  &  J080125.1$-$215956  &   729   &   56   &  3.6  &  17.4  &  0.96  &  2.05   &    8&                                                   \\
HIZSS 035   $ $    &  J080953.9$-$414131  &  1995   &  318   &   24.0  &  16.9  &  5.23  &  0.17   &    6&                                                   \\
HIPASS J0834$-$18  &  J083414.2$-$184215  &  4161   &  147   &  6.7  & 15.11  &  0.34  &  0.82   &    4& [DDR2005]                                         \\
HIPASS J0835$-$58  &  J083513.4$-$583646  &  1050   &   61   &    4.0  &    16.0  &  0.66  &  0.83   &    9& ESO 125-01                                        \\
HIPASS J0836$+$05  &  J083633.8$+$051040  &  1866   &   60   &  2.6  &  17.5  &  0.12  &  3.53   &   10&                                                   \\
HIPASS J0838$-$61  &  J083827.1$-$612330  &  2973   &   91   &  5.9  &  16.5  &  0.83  &  1.66   &    7&                                                   \\
HIPASS J0844$-$33  &  J084355.0$-$330855  &  2316   &  201   & 10.6  &  16.8  &  1.65  &  1.84   &    7&                                       \\
HIPASS J0846$-$76  &  J084624.1$-$765934  &  1520   &   50   &  2.4  &  16.9  &  0.67  &  1.13   &   10&                                                   \\
HIPASS J0848$-$33  &  J084903.4$-$330255  &  1320   &  116   &  4.2  &  17.8  &   1.4  &  2.31   &    7&                                                   \\
HIPASS J0848$-$26  &  J084906.0$-$261918  &   809   &   61   &  3.4  &    15.0  &  0.41  &  0.35   &    5& ESO 496-010                                      \\
HIPASS J0854$-$54  &  J085432.4$-$544645  &  2729   &   59   &  3.8  &  17.1  &  1.51  &  0.99   &    8&                                                   \\
HIPASS J0856$-$11  &  J085651.8$-$110827  &  1879   &   68   &    3.0  &    17.0  &  0.22  &  2.33   &    9&                                                   \\
HIPASS J0857$-$29  &  J085705.6$-$290723  &  1967   &   33   &  3.9  &  17.1  &  0.78  &  2.00   &    8&                                                   \\
HIZSS 049   $ $    &  J085809.4$-$390953  &  1203   &   49   &  3.1  &  17.8  &  2.61  &  0.56   &    7&                                                   \\
HIPASS J0900$-$37  &  J090056.4$-$370723  &  1105   &   97   &  9.1  &  17.8  &  2.52  &  1.79   &    8&                                                   \\
HIPASS J0906$-$34  &  J090656.8$-$341910  &  1410   &   81   &  3.4  &  17.4  &  1.49  &  1.19   &   10&                                                   \\
HIPASS J0916$-$23b &  J091658.0$-$231647  &   837   &   40   &  1.1  & 15.83  &  0.47  &  0.23   &    8& ESO 497-035                                       \\
HIPASS J0917$-$34  &  J091741.7$-$344644  &  2366   &  154   &  8.4  &  17.4  &  1.19  &  3.87   &    8&                       \\
HIPASS J0920$-$34  &  J092044.6$-$342628  &  2410   &  157   &    7.0  &  14.9  &   1.10  &  0.35   &    5& ESO 372-015           \\
HIPASS J0921$-$32  &  J092203.3$-$324717  &  1094   &   97   &  6.9  &  17.2  &   0.80  &  3.80   &    9&                       \\
HIPASS J0926$-$32  &  J092650.1$-$324748  &  1188   &  131   & 10.6  &  16.4  &  0.59  &  3.37   &    7& ESO 373-01            \\
HIPASS J0939$+$00  &  J093857.1$+$004133  &  2076   &   66   &  3.3  &  16.5  &   0.30  &  1.51   &   10& SDSS                  \\
HIPASS J0941$-$15  &  J094059.4$-$153421  &  3619   &  154   &  4.9  &  16.2  &  0.25  &  1.79   &    6&                       \\
HIPASS J0941$-$02  &  J094101.8$-$024303  &  1633   &  128   &  4.5  &  16.9  &  0.15  &  3.40   &    8&                       \\
\end{tabular}
\end{table}

\begin{table}
\begin{tabular}{lrrrrlrrrl} \hline
\hline
\multicolumn{1}{c}{(1)}&
\multicolumn{1}{c}{(2)}&(3)&(4)&(5)&(6)&(7)&(8)&(9)&(10)\\
\hline

HIPASS J0942$+$04  &  J094250.9$+$045324  &  1955   &   66   &  2.9  &    17.0  &  0.24  &  2.21   &    8&                       \\
HIPASS J0944$-$31a &  J094431.8$-$313320  &  2123   &   73   &  3.9  & 16.52  &  0.45  &  1.58   &    9& ESO 434-030           \\
HIZSS 056   $ $    &  J094524.5$-$480828  &   882   &   79   &  7.9  &    17.0  &  2.09  &  1.10   &    8&                       \\
HIPASS J0946$-$15  &  J094622.7$-$154523  &  3164   &   82   &  3.5  &  16.5  &  0.28  &  1.63   &   10&                       \\
HIPASS J0952$-$49  &  J095310.9$-$491950  &  2258   &  100   &    4.0  &    17.0  &  2.08  &  0.56   &    7&                       \\
HIPASS J0958$-$29  &  J095813.6$-$292451  &  2311   &  127   &   14.0  &  15.9  &  0.36  &  3.47   &    5& 6dF                   \\
HIPASS J0958$-$85  &  J100014.0$-$854140  &  1965   &   64   &  2.1  &  16.9  &  0.78  &  0.90   &    8&                       \\
HIPASS J1000$-$77  &  J100108.0$-$771826  &  1716   &  100   &  2.2  &    18.0  &  2.42  &  0.57   &   10&                 \\
HIPASS J1002$-$33  &  J100149.0$-$331542  &  2351   &   98   &    4.0  & 15.67  &   0.40  &  0.78   &    8& ESO 374-021           \\
HIPASS J1003$-$49  &  J100300.7$-$492904  &  2411   &  107   &  7.5  &  17.3  &  2.26  &  1.18   &    7&                       \\
HIPASS J1005$-$28  &  J100539.5$-$282644  &  1037   &  100   &    9.0  & 15.74  &  0.34  &  1.98   &    7& ESO 435-039   \\
HIPASS J1006$-$49  &  J100616.5$-$492829  &  2411   &  140   &  9.2  &    18.0  &  1.83  &  4.09   &    7&                       \\
HIPASS J1007$-$66  &  J100734.7$-$660635  &  1856   &   77   &    5.0  &    19.0  &  0.97  & 12.36   &   10&                       \\
HIPASS J1008$-$33  &  J100829.3$-$330835  &  1645   &   58   &  3.1  &    17.0  &  0.36  &  2.13   &   10&                       \\
HIPASS J1015$-$34  &  J101553.8$-$340653  &  2608   &   27   &  5.4  & 16.17  &  0.38  &  1.69   &    9& ESO 375-03            \\
HIPASS J1016$-$39  &  J101607.2$-$395923  &  3006   &   69   &  8.1  &  16.6  &  0.57  &  3.16   &    8&                       \\
HIPASS J1024$-$12  &  J102428.3$-$122557  &   628   &   45   &  2.6  &  17.9  &   0.30  &  4.33   &   10& KKSG 19               \\
HIPASS J1057$-$21  &  J105709.9$-$213612  &  2075   &   64   &  4.3  &  17.3  &  0.23  &  4.37   &    8&                       \\
HIPASS J1100$-$16  &  J110042.5$-$165436  &  3890   &   59   &  3.8  &  16.2  &  0.23  &  1.41   &    6&                       \\
HIPASS J1117$-$05  &  J111732.9$-$051639  &  3990   &  143   &  4.3  & 16.12  &  0.16  &  1.57   &    6& 2dF         \\
HIPASS J1122$+$13  &  J112313.5$+$134254  &   896   &   27   & 14.1  &  17.5  &  0.12  & 19.05   &   10& ELSB on POSS-IIB \\
HIPASS J1126$-$76  &  J112622.6$-$770012  &  2323   &  159   &  8.6  &  16.5  &  1.52  &  1.28   &    5& 2MASS                 \\
HIPASS J1135$-$46  &  J113532.2$-$464310  &  2827   &  176   &  6.2  &  16.5  &  0.49  &  2.38   &    5& 2MASS                 \\
HIPASS J1152$-$14  &  J115205.1$-$145330  &  2189   &  105   &  4.4  &  16.5  &  0.16  &  2.29   &   10&                       \\
HIPASS J1154$+$12  &  J115412.7$+$122604  &  1005   &   48   &    2.0  &  17.9  &  0.13  &  3.87   &   10&                       \\
HIPASS J1158$-$49  &  J115901.1$-$495102  &  3430   &   84   &  2.1  & 15.17  &  0.53  &  0.23   &    7& [DDR2005]             \\
HIPASS J1201$-$15  &  J120141.0$-$150630  &  1364   &   77   &  4.4  &    15.0  &  0.21  &  0.55   &    9& MCG-02-31-011 \\
HIPASS J1204$-$22  &  J120412.8$-$220635  &  1695   &  130   &  5.1  & 16.39  &  0.22  &  2.27   &    9& NPM1G-21.0322         \\
HIPASS J1215$+$12  &  J121503.9$+$130155  &  2090   &   25   &  2.1  &  16.4  &  0.14  &  1.02   &    8& VCC0132               \\
HIPASS J1215$-$17  &  J121545.1$-$173806  &  1391   &  102   &  5.1  &    16.0  &  0.27  &  1.51   &   10& in a star halo        \\
\end{tabular}
\end{table}

\begin{table}
\begin{tabular}{lrrrrlrrrl} \hline
\hline
\multicolumn{1}{c}{(1)}&
\multicolumn{1}{c}{(2)}&(3)&(4)&(5)&(6)&(7)&(8)&(9)&(10)\\
\hline

HIPASS J1221$-$36  &  J122124.5$-$360638  &  2598   &   47   &  3.2  &  16.8  &  0.35  &  1.84   &    7&                       \\
HIPASS J1225$-$42  &  J122501.6$-$430245  &  2138   &   59   &  3.7  &  14.5  &   0.40  &  0.24   &    5& IRAS                  \\
HIPASS J1227$-$50  &  J122706.8$-$502031  &  3557   &   74   &  2.2  & 15.97  &  0.89  &  0.36   &    7& [DDR2005]             \\
HIPASS J1228$-$50  &  J122858.4$-$500916  &  3694   &  149   &  5.8  &  14.9  &  0.77  &  0.39   &    3& 2MASS                                     \\
HIPASS J1231$+$20  &  J123142.4$+$202853  &  1332   &   69   &  5.1  &  15.6  &  0.12  &  1.20   &    9& CGCG129-006         \\
HIPASS J1233$-$00  &  J123307.9$-$003159  &   729   &   43   &    3.0  & 17.28  &   0.10  &  3.37   &    8& UA285=KDG155 \\
HIPASS J1239$-$04  &  J123901.9$-$043352  &  2412   &   69   &  4.8  & 16.14  &  0.12  &  1.85   &    8& PGC 042298                                                   \\
HIPASS J1239$-$07  &  J123944.7$-$070519  &   929   &   53   &  1.9  & 16.35  &   0.10  &  0.91   &   10& APMUKS                                                       \\
HIPASS J1242$-$77  &  J124206.5$-$775530  &  2685   &   53   &  3.8  &  17.7  &  1.71  &  1.43   &   10&                                                              \\
HIPASS J1244$-$08  &  J124513.3$-$082131  &  2878   &   84   &  7.5  &  15.5  &  0.15  &  1.57   &    8&                                                   \\
HIPASS J1248$-$46  &  J124817.8$-$461304  &  3414   &  279   &  5.8  & 16.28  &  0.41  &  1.96   &    8& ESO 268-G042                                                 \\
HIPASS J1248$-$45b &  J124904.5$-$451120  &  2205   &   78   &  2.4  &    17.0  &  0.41  &  1.57   &    8&                                                              \\
HIPASS J1252$-$31  &  J125235.2$-$315314  &  3549   &  132   &    8.0  & 15.18  &  0.37  &  1.02   &    8& ESO 442-028                                                  \\
HIPASS J1258$-$45  &  J125803.5$-$454830  &  2082   &   70   &  4.3  &  16.4  &  0.39  &  1.64   &    8& KKs52                                                        \\
HIPASS J1303$-$52  &  J130306.6$-$521129  &  1310   &   64   &  4.2  &  17.4  &  1.58  &  1.34   &   10&                                                              \\
HIPASS J1304$-$02  &  J130446.5$-$025217  &  1271   &   78   &  5.6  & 16.37  &  0.11  &  2.70   &   10& 2dFGRS                                                       \\
HIPASS J1305$-$51  &  J130504.2$-$514221  &  4095   &  130   &  4.8  &  17.2  &  1.13  &  1.94   &    7& PGC 045237                                                   \\
HIPASS J1305$-$45  &  J130517.6$-$453925  &  3066   &   53   &  2.7  &  16.9  &  0.39  &  1.64   &    8&                                                              \\
HIPASS J1312$-$15  &  J131226.3$-$154752  &  2953   &  112   &  7.8  & 14.01  &  0.28  &  0.37   &    1& NGC 5010                                                     \\
HIPASS J1314$-$21  &  J131409.4$-$213942  &  2993   &   47   &  2.5  &  16.1  &  0.62  &  0.59   &    7& ESO 576-G015                                                 \\
HIPASS J1315$-$22  &  J131448.2$-$220450  &  1399   &   91   &  4.7  &  17.5  &  0.58  &  4.17   &    8&                                                              \\
HIPASS J1316$-$50  &  J131601.4$-$501620  &  3808   &  314   & 18.3  &  15.4  &  0.81  &  1.91   &    5& ESO 219-43     \\
HIPASS J1320$-$14  &  J132013.0$-$142732  &  2750   &   45   &  4.2  &  16.5  &  0.38  &  1.79   &   10& pec,[MMB2004]                                                \\
HIPASS J1321$-$31  &  J132108.3$-$313145  &   571   &   39   &  5.6  &  17.1  &  0.27  &  4.57   &    9& KK195                                                        \\
HIPASS J1321$-$31b &  J132206.3$-$312442  &  2252   &  195   &    5.0  &  16.6  &  0.27  &  2.58   &    8&                                                         \\
HIPASS J1326$-$09  &  J132551.0$-$093219  &  1803   &   98   &  5.3  &  16.7  &  0.21  &  3.16   &    9&                                                              \\
HIPASS J1328$-$46  &  J132738.0$-$460909  &  1391   &   47   &  4.1  &  17.6  &  0.49  &  4.33   &   10&                                                              \\
HIDEEP J1328$-$28  &  J132823.1$-$281817  &  2053   &        &       & 16.94  &  0.25  &         &   10&                                                              \\
HIDEEP J1333$-$33  &  J133315.4$-$334737  &  2340   &        &       &  16.4  &  0.25  &         &    9&                                                         \\
\end{tabular}
\end{table}

\begin{table}
\begin{tabular}{lrrrrlrrrl} \hline
\hline
\multicolumn{1}{c}{(1)}&
\multicolumn{1}{c}{(2)}&(3)&(4)&(5)&(6)&(7)&(8)&(9)&(10)\\
\hline

HIPASS J1333$-$09  &  J133401.1$-$090900  &  2915   &  149   &  5.9  &  16.1  &  0.17  &  2.11   &    8&                                                              \\
HIPASS J1334$-$12  &  J133439.7$-$121950  &  1511   &   76   &  2.2  &    15.0  &  0.24  &  0.27   &    8& KDG227                                                       \\
HIDEEP J1336$-$33  &  J133700.0$-$332147  &   591   &        &       &  17.3  &  0.26  &         &   10&                                                              \\
HIDEEP J1337$-$31  &  J133658.9$-$311905  &  1693   &        &       &  18.8  &  0.26  &         &    5& 6dF                                                          \\
HIDEEP J1337$-$28  &  J133718.2$-$283929  &  2361   &        &       &  15.5  &  0.25  &         &    4& IC 4303                                                      \\
HIPASS J1338$-$30  &  J133808.0$-$305506  &  1667   &  135   &  6.4  &    16.0  &  0.24  &  1.96   &    9& bluish,2MASS,diffuse tail(?!)                     \\
HIDEEP J1338$-$30  &  J133845.3$-$305406  &  3650   &        &       & 15.93  &  0.24  &         &    4&ESO 444-090                                                   \\
HIPASS J1340$-$31  &  J134020.9$-$314204  &  1658   &   96   &  5.2  & 16.57  &  0.24  &  2.68   &    8& ESO 445-007                                                  \\
HIPASS J1342$-$48a &  J134200.4$-$481837  &  1372   &  114   &  7.1  &  14.4  &   0.40  &  0.43   &    7& ESO 220-032                                                  \\
HIPASS J1342$-$19  &  J134246.9$-$193454  &  1410   &   54   &  2.7  & 16.42  &  0.47  &  0.98   &   10& ESO 577-27                                                   \\
HIPASS J1345$-$35  &  J134455.4$-$352450  &  1018   &   52   &  2.6  & 16.01  &  0.25  &  0.79   &   10& ESO 383-70                                                   \\
HIPASS J1348$-$37  &  J134833.5$-$375805  &   501   &   39   &  2.5  &    18.0  &  0.33  &  4.41   &   10&                                                              \\
HIPASS J1350$-$37  &  J135033.3$-$371831  &  1077   &  146   &   11.0  &  14.3  &  0.33  &  0.64   &    7& ESO 383-091                                                  \\
HIPASS J1351$-$47  &  J135118.9$-$465825  &   528   &   38   &  3.5  &  17.5  &  0.62  &  2.99   &   10&                                                              \\
HIPASS J1354$-$49  &  J135407.2$-$492750  &  1788   &  103   &  5.1  &    17.0  &  0.83  &  2.27   &    8&                                             \\
HIPASS J1357$-$46  &  J135739.2$-$462618  &  2850   &  126   &  3.7  &  16.9  &  0.55  &  1.94   &    8&                                             \\
HIPASS J1400$-$06  &  J135958.8$-$062808  &  2604   &   99   &  3.2  & 16.18  &  0.14  &  1.26   &    9& APMUKS                                      \\
HIPASS J1401$-$05  &  J140131.8$-$054933  &  1564   &  128   &  4.3  & 16.49  &  0.12  &  2.29   &    7& APMUKS                                      \\
HIPASS J1402$-$48  &  J140312.0$-$483739  &  2807   &  142   &    5.0  &  16.4  &  0.81  &  1.31   &    7&                                             \\
HIPASS J1405$-$51  &  J140539.9$-$510445  &  4092   &  111   &  9.7  &  17.2  &  1.53  &  2.73   &    8&                                             \\
HIPASS J1407$-$75  &  J140729.2$-$755314  &  2821   &  132   &  7.7  &  16.8  &  0.72  &  3.16   &    8&                                             \\
HIPASS J1409$-$14  &  J140941.2$-$144620  &  2328   &  106   &  3.7  &  16.4  &  0.35  &  1.47   &    7&                                             \\
HIPASS J1424$-$34  &  J142352.1$-$341529  &  3118   &   45   &  2.4  & 16.82  &  0.53  &  1.19   &    8& ESO 385-020         \\
HIPASS J1425$-$39  &  J142549.0$-$391603  &  1174   &   57   &  4.5  &    17.0  &  0.48  &  2.75   &   10&                                             \\
HIPASS J1428$-$08  &  J142826.8$-$085524  &  1541   &   77   &  5.4  &  16.9  &  0.24  &  3.77   &   10& KKSG46                                      \\
HIPASS J1433$+$01  &  J143353.5$+$012912  &  1827   &   66   &  3.2  &  17.4  &   0.20  &  3.66   &   10&                                             \\
HIPASS J1435$-$17  &  J143525.4$-$171001  &  1574   &  113   &  9.1  &  17.2  &  0.32  &  7.80   &   10& KKSG47                                      \\
HIPASS J1444$-$17  &  J144457.4$-$171453  &  2826   &  105   &  5.5  &  16.6  &   0.50  &  2.29   &    8&                                             \\
HIPASS J1452$-$50  &  J145256.6$-$505352  &  1421   &  122   &  5.9  &  17.5  &  2.25  &  1.13   &    7&  ESO 223-03                          \\
HIPASS J1546$-$32  &  J154626.3$-$323745  &  2231   &   63   &  5.6  & 16.53  &  0.46  &  2.27   &   10& ESO 450-12                                  \\
HIPASS J1548$+$16  &  J154902.8$+$164249  &  2052   &   94   &  5.3  &    18.0  &  0.16  & 10.96   &    8&                                             \\
\end{tabular}
\end{table}

\begin{table}
\begin{tabular}{lrrrrlrrrl} \hline
\hline
\multicolumn{1}{c}{(1)}&
\multicolumn{1}{c}{(2)}&(3)&(4)&(5)&(6)&(7)&(8)&(9)&(10)\\
\hline

HIPASS J1551$-$70  &  J155018.5$-$702854  &  3328   &  140   &  5.6  &    16.0  &  0.61  &  1.21   &    6& 2MASX                               \\
HIPASS J1604$-$75  &  J160302.5$-$753932  &  2912   &  140   &  3.1  &  16.5  &  0.28  &  1.45   &    8& ESO 042-016                        \\
HIPASS J1605$-$04  &  J160540.8$-$043420  &  1611   &   91   &  7.2  &  16.2  &  1.19  &  1.10   &   10& KKSG48                                      \\
HIPASS J1615$-$17  &  J161545.5$-$175028  &  2371   &  197   &    8.0  &    16.0  &  1.51  &  0.76   &    8& VLSB,2'x1',cirrus-like                     \\
HIPASS J1700$-$12  &  J170006.3$-$120001  &  1298   &   77   &  5.7  &  17.3  &  1.23  &  2.31   &    8&                                             \\
HIPASS J1700$-$07  &  J170008.1$-$072324  &  1558   &   56   &  4.9  &  17.7  &  2.38  &  1.00   &   10&                                             \\
HIPASS J1711$-$47  &  J171137.0$-$473604  &  2182   &  219   & 20.1  &  17.6  &  2.97  &  2.17   &    6&                                             \\
HIPASS J1730$-$60  &  J173100.2$-$601138  &  1565   &   46   &  3.3  &  17.4  &  0.37  &  3.25   &    8&                                             \\
HIPASS J1738$-$57  &  J173842.9$-$571525  &   858   &   75   &  4.5  &  17.5  &  0.41  &  4.66   &   10&                                             \\
HIPASS J1739$-$51  &  J173910.1$-$510457  &  3802   &  146   &   10.0  &  17.8  &  0.71  & 10.38   &    8&                                             \\
HIPASS J1750$+$21  &  J175011.5$+$211556  &  3222   &   82   &  2.2  &  16.4  &   0.40  &  0.84   &    6& 2MASS                                       \\
HIPASS J1752$-$59  &  J175250.9$-$594057  &  2736   &   46   &  3.3  &  17.7  &   0.40  &  4.17   &   10&                                             \\
HIPASS J1758$+$14  &  J175839.5$+$144658  &  2957   &  134   &  4.1  &  16.8  &  0.52  &  2.01   &    6&                                             \\
HIPASS J1809$-$05  &  J180955.0$-$055423  &  2935   &  250   &  6.6  &  18.8  &  4.94  &  0.35   &    7& 2MASS                                       \\
HIPASS J1810$-$01  &  J181042.1$-$010427  &  2092   &  165   &  5.8  &  18.5  &  4.87  &  0.25   &    7&                                             \\
HIPASS J1812$-$74  &  J181220.0$-$744957  &  3216   &   76   &  3.8  &  17.8  &   0.60  &  4.37   &   10& in a star halo                              \\
HIPASS J1819$+$01  &  J181923.8$+$011042  &  2571   &  114   & 12.8  &  18.2  &  4.35  &  0.67   &    6&                                             \\
HIPASS J1824$-$01  &  J182500.2$-$012833  &  2875   &  351   & 30.5  &    19.0  &  6.95  &  0.30   &    4&                                             \\
HIPASS J1832$+$06  &  J183214.6$+$062530  &  2823   &   91   &    7.0  &  17.7  &  2.06  &  1.91   &    7&                                             \\
HIPASS J1859$-$57  &  J185950.4$-$570126  &  3461   &   42   &  2.3  &  16.4  &  0.32  &  0.94   &    6&                                             \\
HIPASS J1915$-$57  &  J191535.4$-$570559  &  2991   &   61   &  2.5  &  16.8  &   0.30  &  1.50   &    7&                                             \\
HIPASS J1919$-$33  &  J191932.8$-$334603  &  2078   &  100   &  3.1  &    17.0  &  0.38  &  2.09   &    7&                                             \\
HIPASS J1926$-$74  &  J192729.4$-$740454  &  2624   &   43   &  2.3  &  17.2  &  0.34  &  1.92   &    9&                                             \\
HIPASS J1932$-$55  &  J193216.5$-$555429  &  3413   &  203   &   13.0  & 13.39  &  0.27  &  0.35   &    0& NGC 6799                                    \\
HIPASS J1937$-$52  &  J193738.5$-$520433  &  3157   &  199   & 21.5  &  14.5  &  0.24  &  1.64   &    9& IC 4875                                     \\
HIPASS J1937$+$09  &  J193744.4$+$092125  &  3148   &   81   &  3.3  &  17.5  &  1.92  &  0.86   &    7&                                             \\
HIPASS J1951$+$01  &  J195157.3$+$013140  &  1275   &   87   &  3.9  &  17.7  &  1.07  &  2.65   &    8&                                             \\
HIPASS J2001$-$47  &  J200156.5$-$471550  &  3490   &   90   &  3.3  & 15.74  &  0.18  &  0.84   &    8& ESO 284-010                                 \\
HIPASS J2011$-$09  &  J201059.2$-$095545  &  2868   &   86   &  5.6  &  17.4  &  0.39  &  5.40   &    8& 6dF                                         \\
HIPASS J2010$-$11  &  J201047.6$-$113829  &  3204   &  100   &  6.9  &    16.0  &  0.56  &  1.57   &    7&                                             \\
\end{tabular}
\end{table}

\begin{table}
\begin{tabular}{lrrrrlrrrl} \hline
\hline
\multicolumn{1}{c}{(1)}&
\multicolumn{1}{c}{(2)}&(3)&(4)&(5)&(6)&(7)&(8)&(9)&(10)\\
\hline
HIPASS J2015$+$12  &  J201548.6$+$124127  &  1951   &   52   &  4.6  &  18.3  &  0.87  &  6.55   &   10&                                             \\
HIPASS J2041$-$16  &  J204118.7$-$160424  &  3452   &   99   &  6.4  &  16.3  &   0.20  &  2.68   &    5& [DDR2005]                                   \\
HIPASS J2045$-$53  &  J204513.7$-$534353  &  3255   &  112   &  5.8  & 15.89  &  0.21  &  1.64   &    9& APM                                         \\
HIPASS J2100$-$55  &  J205956.9$-$553343  &  2033   &  188   &  8.8  & 13.96  &  0.21  &  0.42   &    2& NGC 6990                                    \\
HIPASS J2111$-$50  &  J211156.8$-$502111  &  2790   &   86   &  4.1  &  16.2  &  0.11  &  1.69   &    8& ESO 235-073                                 \\
HIPASS J2138$-$54  &  J213759.5$-$543130  &  2924   &   96   &  5.7  &  16.9  &  0.09  &  4.57   &    8& APM, blue,in a group?                       \\
HIPASS J2149$-$35  &  J215010.9$-$354228  &  2458   &  104   &  6.5  & 16.89  &   0.10  &  5.11   &   10& ESO 403-036                                 \\
HIPASS J2150$-$23  &  J215050.5$-$230848  &  2342   &  156   &  5.2  &    17.0  &  0.17  &  4.25   &    9& 2MASX                                       \\
HIPASS J2156$-$49  &  J215556.9$-$491438  &  1865   &   91   &  3.9  & 17.34  &   0.10  &  4.66   &    8& APMUKS                                      \\
HIPASS J2214$-$67  &  J221513.9$-$672636  &  3232   &   80   &  1.7  &  15.4  &  0.14  &  0.33   &    4& FAIRALL1005                         \\
HIPASS J2217$-$42  &  J221654.8$-$424512  &  2198   &   49   &  2.9  &    17.0  &  0.06  &  2.63   &   10& AM                                          \\
HIPASS J2229$-$79  &  J223020.5$-$794131  &  2531   &  154   &  8.3  & 15.06  &  0.55  &  0.80   &    7& ESO 027-012+companion                       \\
HIPASS J2250$+$00  &  J225024.4$+$005241  &  1692   &  142   &  2.8  &  17.2  &  0.42  &  2.19   &   10& APM                                         \\
HIPASS J2251$-$20b &  J225148.7$-$203629  &  3145   &  328   & 14.8  & 12.63  &  0.14  &  0.22   &    2& NGC 7392                                    \\
HIPASS J2307$-$61  &  J230720.8$-$614050  &  3223   &   48   &  2.6  &  17.5  &  0.09  &  3.63   &   10& VLSB                                        \\
HIPASS J2308$+$17  &  J230851.4$+$171236  &  1764   &  142   & 16.7  &  15.8  &  0.56  &  3.16   &    8&                                             \\
HIPASS J2311$-$42  &  J231110.9$-$425051  &  1381   &   94   &  3.4  & 15.72  &  0.04  &  0.96   &    8& ESO 291-003                                 \\
HIPASS J2323$-$50  &  J232359.4$-$503127  &  1579   &  131   &  5.1  & 14.99  &  0.05  &  0.73   &    1& ESO 240-02                                  \\
HIPASS J2340$-$20  &  J234032.6$-$205639  &  1397   &  114   &  5.4  & 16.51  &  0.11  &  2.96   &    7& ESO 606-01+companion                        \\
HIPASS J2344$-$07  &  J234429.7$-$073547  &  1832   &   76   &  5.7  & 16.05  &  0.14  &  2.00   &    8&                                             \\
HIPASS J2358$+$04  &  J235807.3$+$044905  &  3035   &  111   &  4.2  &  17.9  &  0.21  &  7.59   &    8&                                             \\
HIPASS J2359$+$02  &  J235913.4$+$024324  &  2616   &  148   &  4.6  &  16.4  &  0.12  &  2.27   &    6&                                             \\
\hline
\end{tabular}
\end{table}